\documentclass[twocolumn,showpacs,amssymb,prd]{revtex4}

\usepackage{graphicx}
\usepackage{dcolumn}
\usepackage{bm}

\def\beq{\begin{equation}}
\def\enq{\end{equation}}
\def\ba{\begin{eqnarray}}
\def\ea{\end{eqnarray}}
\def\Mesz{M\'esz\'aros~}
\def\<{<\!\!}
\def\>{\!\!>}
\def\ra{\rightarrow}

\def\vareps{\varepsilon}

\begin{document}
\input{epsf}

\title{Neutrino Tomography of Gamma Ray Bursts and Massive Stellar
Collapses}

\author{Soebur Razzaque,$^1$ Peter M\'esz\'aros$^1$ and Eli
Waxman$^2$}

\affiliation{$^1$Dpt Astronomy \& Astrophysics, Dpt Physics,
 Pennsylvania State Univ., University Park, PA 16802, USA \\
 $^2$Department of Condensed Matter Physics, Weizmann Institute of
 Science, Rehovot 76100, Israel}

\begin{abstract}
Neutrinos at energies above TeV can serve as probes of the stellar
progenitor and jet dynamics of gamma ray bursts arising from stellar
core collapses. They can also probe collapses which do not lead to
gamma-rays, which may be much more numerous. We calculate detailed
neutrino spectra from shock accelerated protons in jets just below the
outer stellar envelope, before their emergence. We present neutrino
flux estimates from such pre-burst jets for two different massive
stellar progenitor models. These should be distinguishable by IceCube,
and we discuss the implications.
\end{abstract}

\date{\today}
\pacs{96.40.Tv,98.70.Rz,98.70.Sa}

\maketitle

\section{Introduction}

The gamma-ray bursts (GRB) which have so far been accurately localized
are associated with regions of active star formation, and their
progenitors are thought to be massive stars. The leading model for
such bursts involves a relativistic jet, produced following the
collapse of the core of the massive stellar progenitor
\cite{woo03}. In this model the $\gamma$-rays are produced by
synchrotron or inverse Compton radiation from Fermi accelerated
electrons in optically thin shocks (see \cite{mesz02} for a review),
after the jet has emerged from the stellar envelope. The same
optically thin shocks should accelerate relativistic protons
\cite{wax95}, and lead to $\sim 100$ TeV neutrinos via interactions
with the observed MeV $\gamma$-rays \cite{wb97}. However, while the
jets are still inside the star, shock-accelerated protons can produce
$\sim$ TeV neutrinos through photomeson interactions with thermal
X-rays in the sub-stellar jet cavity \cite{mw01}.

In this paper we discuss a more general class of massive stellar
collapses, in which jet formation may be ubiquitous, but not all of
which emerge to be associated with detectable GRBs.  Before their
successful or failed emergence from the star, the jets can accelerate
protons which undergo a more complex sequence of high energy
interactions than previously realized. These depend not only on the
jet and central engine characteristics but also on the location of the
shocks and on the outer dimensions of the stellar progenitor, thus
providing potentially useful diagnostics for the type of progenitor as
well as the jet and shock parameters. Protons accelerated in
sub-stellar jet shocks first undergo photomeson interactions with
thermalized shock photons, as well as $pp,~pn$ interactions with
thermal nucleons in the jet frame.  This modifies the relativistic
proton spectrum reaching the end of the jet cavity, where the protons
undergo a second set of photomeson interactions with stellar X-ray
photons and $pp,~pn$ interactions with cold nucleons in the stellar
frame. The fraction of collapses producing jets which subsequently
emerge from the star to produce electromagnetically detectable GRBs
are expected to be preceded by a precursor neutrino signal at energies
$\gtrsim$ TeV, which is significantly different from the previously
calculated $\gtrsim 100$ TeV neutrino signals coincident with the
$\gamma$-rays \cite{wb97}.  The fraction of stellar collapses leading
to jets which do not emerge would have similar neutrino signals, but
they could be more numerous and hence their diffuse flux could be more
important.

We discuss our jet models in Sec. \ref{sec:jetmodel}, proton and
electron acceleration in the internal shocks in
Sec. \ref{sec:protonacc} and proton interactions in
Sec. \ref{sec:protonint}.  We discuss neutrino production mechanisms
in Sec. \ref{sec:nuproduction} and calculate observed neutrino flux in
Sec. \ref{sec:nuflux}.  We summarize and discuss implications of our
results in Sec. \ref{sec:implications}.

\section{Collapsar and jet models}
\label{sec:jetmodel}

We take a simplified model for the jet and progenitor star, similar to
that described in Ref. \cite{mw01}. The GRB progenitor is taken to be
a massive star with a He core and H envelope. As an example, the
parameters chosen are a core radius $r_{\rm He} \approx 10^{11.5}$ cm
, where the density is $\rho_{\rm He} \sim 10^{-3} \rho_{-3}$
g/cm$^3$.  We take two cases, one case (H) where the core is
surrounded by an H envelope of size $r_* \sim 10^{13}$ cm, where the
density $\rho_{\rm H} \approx 10^{-7} \rho_{-7}$ g/cm$^3$, and another
case (He) where a similar He core has lost its surrounding H envelope.
Numerical simulations of collapsar models leading to black hole
formation (e.g. \cite{woo03} indicate that a relativistic jet can be
launched along the stellar rotation axis, powered either by thermal
$\nu{\bar\nu}$ annihilation or MHD stresses coupling to the black hole
and to the debris disk falling back onto the hole). The jet life time
is limited by the gas fall-back time onto the black hole, and should
be comparable to GRB durations, $\lesssim 10^2$~s. Variability in the
efective jet Lorentz factor arises both from variability in the black
hole accretion rate, and from instabilities at the jet-stellar
interface on the way out from the star.

Internal shocks occur at radii $r_{sh}\sim c \delta t \Gamma^2$ in the
highly variable jet outflow, where we expect fluctuations in the bulk
Lorentz factor $\Gamma \sim 10^2-10^3$ over a wide range of time
scales, $0.1 {\rm ms}\lesssim \delta t \lesssim 1$ s. The observed
$\gamma$-rays are produced in jet internal shocks between regions with
high enough $\Gamma$, $\delta t$, which occur in the optically thin
environment outside the progenitor star
(e.g. \cite{mesz02}). Fluctuations with lower $\Gamma$, $\delta t$ are
also likely to produce internal shocks at smaller radii below the
stellar surface \cite{mw01}, which will be optically thick.  In
observed $\gamma$-ray light curves, variability on sub-ms timescales
is present, the power at high frequencies \cite{belob00} being subject
to Poisson noise uncertainties due to low photon counts.  In
sub-stellar shocks with smaller lengthscales and Lorentz factors the
high frequency power could be relatively more important.  Such
episodes of lower $\Gamma$ or shorter variability could arise from
instabilities leading to the up and down variation of $\Gamma$ during
propagation of the jet inside the star, as shown in numerical
simulations \cite{woo03}.

Here we are interested mainly in these opaque, sub-surface internal
shocks.  We consider the jets at a time when they are still inside the
star, but have advanced to within a factor 1/few of the outer stellar
surface. As specific numerical examples we take the internal shock
radii at $r_{sh,{\rm H}} \sim \alpha 10^{12.5}r_{12.5}$ cm and $r_{sh,
{\rm He}} \sim \alpha 10^{11}r_{11}$ cm, where $\alpha \lesssim 0.5$,
and the jet termination radii are taken to be $r_{jet,{\rm H}} \sim
10^{12.5}r_{12.5}$ cm and $r_{jet,{\rm He}} \sim 10^{11}r_{11}$ cm for
the two types of progenitors. The internal shocks at $r_{sh}$ provide
a particle accelerator sufficiently far inside the jet cavity, where
they are not subject to the large radiative losses in the high photon
and nucleon density which are present further out near the jet
termination radius $r_{jet}$, i.e., we consider $r_{sh} < r_{jet} <
r_*$. The jets are optically thick to Thomson scattering and the
density of photons from the jet termination surface drops off very
fast with distance measured inwards into the jet \cite{mw01}.  The
shocks in the sub-stellar shocks are likely to be collisionless, since
the typical collisionless width $c/\omega_{p,i}$ (where $\omega_{p,i}$
is the ion plasma frequency) is much smaller than the radiation
scattering and the particle collision mean free paths.  We introduce a
parameter $\vareps_{\rm op}$ that characterizes the fraction of jet
energy that is dissipated in the sub-stellar shocks (this energy is
given back to the jet, due to the high optical depth, except for the
fraction that escapes as neutrino energy). The neutrino fluxes we
derive will be proportional to $\vareps_{\rm op}$.  We cannot estimate
$\vareps_{\rm op}$ directly, since we do not see these
collisions. However, it is likely to be significant due to the
fluctuations in jet parameters, and we will assume $\vareps_{\rm op}
\sim 1/2$.

For observed isotropic GRB luminosities $L_{\gamma}^{iso} = 10^{52}
L_{52}$ ergs/s at the termination radius of a sub-surface jet, the
jet-frame proton number density is
\ba 
n_{p, jet} &=& L_{\gamma}^{iso}/(4\pi r_{jet}^2 \Gamma_{jet}^2 m_p
c^3) \nonumber \\ & \approx & \cases{ 1.8 \times 10^{14}L_{52}
r_{12.5}^{-2} \Gamma_{100}^{-2} \, {\rm cm}^{-3} & \mbox{(H)} \cr 1.8
\times 10^{17}L_{52}r_{11}^{-2} \Gamma_{100}^{-2} \, {\rm cm}^{-3} &
\mbox{(He)}. }
\label{proton-jetdensity} 
\ea
Equating the pressures behind the forward and reverse shocks one finds
that the jet head moves at mildly relativistic or subrelativistic
speeds with Lorentz factor \cite{Waxman:2003uu}
\ba
\Gamma_{h} \approx \cases{ 1.6 L_{52}^{1/4} r_{12.5}^{-1/2}
\rho_{-7}^{-1/4} & \mbox{(H)} \cr 1 L_{52}^{1/4} r_{11}^{-1/2}
\rho_{-3}^{-1/4} & \mbox{(He)}. }
\label{head-lorentz}
\ea
The number density of protons in the stellar plasma frame is $n_p =
2\Gamma_{jet}n_{p,jet}/\Gamma_{h}$.

The termination shock where the jet head impacts the star heats up the
stellar plasma.  The temperature of thermalized shocked photons in the
jet head frame, from $L_{\gamma}^{iso} \approx 4\pi r_{jet}^2 c
\Gamma_{h}^2a_B T^4$, is
\ba
T_{h} \approx \cases{ 1.5 L_{52}^{1/4} r_{12.5}^{-1/2}
\Gamma_{h,1.6}^{-1/2} \, {\rm keV} & \mbox{(H)} \cr 11 L_{52}^{1/4}
r_{11}^{-1/2} \Gamma_{h,1}^{-1/2} \, {\rm keV} & \mbox{(He)}. }
\label{shgamma-temp}
\ea
The corresponding number density of these photons in the jet head
frame, which is approximately the same in the stellar plasma and in
the observer's frame, is
\ba
n_{\gamma,h} &=& L_{\gamma}^{iso}/(4\pi r_{jet}^2 \Gamma_{h}^2 c
T_{h}) \nonumber \\ & \approx & \cases{ 4.3 \times 10^{23}
L_{52}^{3/4} r_{12.5}^{-3/2} \Gamma_{h,1.6}^{-3/2} \, {\rm cm}^{-3} &
\mbox{(H)} \cr 1.9 \times 10^{26} L_{52}^{3/4} r_{11}^{-3/2}
\Gamma_{h,1}^{-3/2} \, {\rm cm}^{-3} & \mbox{(He)}. }
\label{photon-headdensity}
\ea
Photons from the shocked stellar plasma diffuse into the jet through
Thomson scatterings. Their penetration depth (mean-free path) in the
shocked plasma frame is
\ba 
l_{\gamma} &=& \Gamma_{h}/(2\Gamma_{jet} n_{p,jet} \sigma_{\rm
Th}) \nonumber \\ &\approx &\cases{ 6.8 \times 10^7 L_{52}^{-1}
r_{12.5}^{2} \Gamma_{h,1.6} \Gamma_{100} \, {\rm cm} & \mbox{(H)} \cr
3.8 \times 10^4 L_{52}^{-1} r_{11}^{2} \Gamma_{h,1} \Gamma_{100} \,
{\rm cm} & \mbox{(He)}. }
\label{photon-depth}
\ea
The density of diffused photons inside the jet head drops off as
\ba
n_{\gamma}(x) \approx n_{\gamma,h}e^{-x/l_{\gamma}}
\label{sh-gam-dist}
\ea
where $x$ is the distance from the jet head radius $r_{jet}$
\cite{mw01}.

\section{Particle acceleration}
\label{sec:protonacc}

We calculate proton acceleration at the internal shocks which take
place at a radius smaller than the termination shock radius ($r_{sh} <
r_{jet}$). In principle protons can be accelerated at the termination
(forward and reverse) shock too, however, inverse Compton losses
significantly reduce the maximum proton energy there \cite{mw01}.

The maximum energy up to which protons are accelerated in the internal
shocks is limited by synchrotron losses in the jet magnetic field.
(Inverse Compton losses are in the Klein-Nishina limit, [Eq.
(\ref{proton-IC})], and less important than synchrotron).  We estimate
the magnetic field $B_{jet}$ in the internal shocks, in the jet frame,
from the relation $4\pi \alpha^2 r_{sh}^2 c \Gamma_{jet}^2 B_{jet}^2 /
8\pi = \xi_B L_{\gamma}^{iso}$.  Dropping $\alpha \sim 1$ for
notational simplicity and assuming a magnetic field equipartition
fraction $\xi_B = 10^{-1} \xi_{-1}$, we get
\ba
B_{jet} \approx \cases{ 8.2 \times 10^5
\xi_{-1}^{1/2}L_{52}^{1/2}r_{12.5}^{-1} \Gamma_{100}^{-1} \, \mbox{G}
& \mbox{(H)} \cr 2.5 \times 10^7 \xi_{-1}^{1/2}L_{52}^{1/2}r_{11}^{-1}
\Gamma_{100}^{-1} \, \mbox{G} & \mbox{(He)}. }
\label{Bjet-field}
\ea
Equating the synchrotron cooling time
\ba 
t_{p,jet}^{syn} &=& 6\pi m_p^3/(\sigma_{\rm Th} m_e^2
\gamma_{p,jet}B^2_{jet}) \nonumber \\ &\approx & \cases{ 7 \times
10^{6}\gamma_{p,jet}^{-1} \xi_{-1}^{-1}L_{52}^{-1} r_{12.5}^2
\Gamma_{100}^2 \,{\rm s} & \mbox{(H)} \cr 7 \times
10^{3}\gamma_{p,jet}^{-1} \xi_{-1}^{-1}L_{52}^{-1} r_{11}^2
\Gamma_{100}^2 \,{\rm s} & \mbox{(He)}}
\label{sync-time}
\ea
to the acceleration time
\ba
t_{p,jet}^{acc} &\approx & 10 (m_p\gamma_{p,jet}/eB_{jet}) \nonumber
\\ &\approx & \cases{ 1.5 \times 10^{-8} \gamma_{p,jet}
\xi_{-1}^{-1/2}L_{52}^{-1/2} r_{12.5} \Gamma_{100} \,{\rm s} &
\mbox{(H)} \cr 4.7 \times 10^{-10} \gamma_{p,jet}
\xi_{-1}^{-1/2}L_{52}^{-1/2} r_{11} \Gamma_{100} \,{\rm s} &
\mbox{(He),} }
\label{acc-time}
\ea
(both evaluated in the jet frame) we get the maximum proton energy in
the observer's frame as
\ba
E_{p}^{max} &=& m_p \gamma_{p,jet}^{max} \Gamma_{jet} \nonumber \\
&\approx & \cases{ 2.0 \times 10^{9}
\xi_{-1}^{-1/4}L_{52}^{-1/4}r_{12.5}^{1/2} \Gamma_{100}^{5/2} \, {\rm
GeV} & \mbox{(H)} \cr 3.6 \times 10^{8}
\xi_{-1}^{-1/4}L_{52}^{-1/4}r_{11}^{1/2} \Gamma_{100}^{5/2} \, {\rm
GeV} & \mbox{(He)}. }
\label{maxproton-energy}
\ea
This is smaller than maximum proton energy achievable in the observed
optically thin GRBs, since the internal shocks take place at larger
radii in the latter case.  

Electrons are also accelerated in the internal shocks, and the average
electron energy in the jet frame is $\< \gamma_e \>_{jet} = (m_p/m_e)
\vareps_e \approx 3.7 \times 10^2 \vareps_{0.2}$, where $\vareps_e =
0.2 \vareps_{0.2}$ is the fraction of thermal energy which goes into
electrons.  The corresponding peak synchrotron photon energy in the
jet frame is $(3/4)(B_{jet}/B_Q) \< \gamma_e \>^2_{jet} m_e$, where
$B_Q = 4.4 \times 10^{13}$ G.  However, the jet is optically thick to
Thomson scattering, with an optical depth in the jet frame, from
Eq. (\ref{proton-jetdensity}), of
\ba
\tau_{{\rm Th}, jet} &=& \frac{\sigma_{\rm Th} n_{p, jet} r_{sh}}
{\Gamma_{jet}} \approx \cases{
4 \, L_{52}r_{12.5}^{-1}\Gamma_{100}^{-3} & \mbox{(H)} \cr
120 \, L_{52}r_{11}^{-1}\Gamma_{100}^{-3} & \mbox{(He)}. }
\label{thom-syn-opt}
\ea
Note that these values for sub-stellar jets are larger than those in
the internal shocks responsible for $\gamma$-rays outside the star,
due to the smaller distances and higher densities.  Hence the
synchrotron photons thermalize to an approximate black-body
temperature of
\ba 
E_{\gamma,jet}^{syn} &=& \left[ L_{\gamma}^{iso}/(4\pi r_{sh}^2
\Gamma_{jet}^2 c a_{B}) \right]^{1/4} \nonumber \\ &\approx & \cases{
0.2\,L_{52}^{1/4}r_{12.5}^{-1/2}\Gamma_{100}^{-1/2}\, \mbox{KeV} &
\mbox{(H)} \cr 1.2\,L_{52}^{1/4}r_{11}^{-1/2} \Gamma_{100}^{-1/2}\,
\mbox{KeV} & \mbox{(He)} }
\label{syn-phot-E}
\ea
in the jet frame.  The corresponding number density of synchrotron
photons in the jet frame is
\ba
n_{\gamma, jet}^{syn} &=&
L_{\gamma}^{iso}/(4\pi r_{sh}^2 \Gamma^2_{jet} c E_{\gamma, jet}^{syn})
\nonumber \\ &\approx & \cases{
8.0 \times 10^{20}L_{52}^{3/4}r_{12.5}^{-3/2} \Gamma_{100}^{-3/2} \, 
\mbox{cm}^{-3} & \mbox{(H)} \cr
1.4 \times 10^{23}L_{52}^{3/4}r_{11}^{-3/2} \Gamma_{100}^{-3/2} \, 
\mbox{cm}^{-3} & \mbox{(He).} }
\label{syn-phot-density}
\ea

We have also calculated energy losses by the protons due to photopair
and photomeson interactions with synchrotron photons, which could
possibly limit their acceleration.  The Bethe-Heitler process $p\gamma
\ra p e^+e^-$ has a very large cross-section of $\approx 5 \times
10^{-25}$ cm$^2$ at threshold, but the energy lost by the proton is
negligible (although they might play a role in more accurate
calculations of the asymptotic energy losses).  Photomeson
interactions affect proton acceleration at energies only above
$\Delta$ production threshold, which we discuss in the following
section.  The typical jet crossing time in the comoving frame: $30/100
\sim 0.3$ s, for bulk Lorentz factor $\Gamma = 100$, is much longer
than the acceleration time in Eq. (\ref{acc-time}) for all but the
highest proton energies achievable.  Hence the jet crossing time does
not introduce any additional constraints for protons to accelerate to
the maximum observer-frame energies of Eq.  (\ref{maxproton-energy}),
leading to a similar upper limit.

\section{Proton interactions}
\label{sec:protonint}

High energy protons accelerated in the sub-stellar internal shocks
undergo $p\gamma$ and $pp$ interactions both inside the jet and near
its termination boundary with the stellar envelope. We discuss each
case below.

Protons first interact inside the jet with synchrotron photons from
co-accelerated electrons, and with the fraction of jet protons which
are not accelerated at the internal shocks. The dominant inelastic
channel is $p\gamma \ra \Delta$ with $\approx 5 \times 10^{-28}$
cm$^2$ cross-section ($\sigma_{p\gamma \ra \Delta}$) and $\approx$
20\% inelasticity. The $p\gamma$ optical depth at the $\Delta$
resonance from Eq. (\ref{syn-phot-density}) in the jet frame is then
\ba
\tau_{p\gamma, jet}^{syn} &=& (\sigma_{p\gamma \ra \Delta}) n_{\gamma,
jet}^{syn} r_{sh}/\Gamma_{jet} \nonumber \\ & \approx & \cases{ 1.3
\times 10^4 L_{52}^{3/4}r_{12.5}^{-1/2}\Gamma_{100}^{-5/2} &
\mbox{(H)} \cr 7.0 \times 10^4
L_{52}^{3/4}r_{11}^{-1/2}\Gamma_{100}^{-5/2} & \mbox{(He);} }
\label{pg-syn-opt}
\ea
very large in both cases (see e.g. \cite{dermer02} for photomeson
interactions in a different, optically thin, GRB acceleration
scenario).  With an optical depth so large, high energy protons lose
all their energies through $p\gamma$ interactions.  Secondary neutrons
produced in the subprocess $p\gamma \ra \Delta \ra n\pi^{+}$ also
undergo photo-meson processes just like protons.  This scenario is
different from that of optically thin shocks in observed GRBs, where
the $p\gamma$ optical depth is much smaller because of larger shock
radii \cite{wb97,wb99}. The threshold proton energy for $p\gamma$
interaction at the $\Delta$ resonance, from the condition
$E_p^{th}E_{\gamma, jet}^{syn} \approx 0.3 \Gamma_{jet}$ GeV$^2$, in
the observer's frame, from Eq. (\ref{syn-phot-E}), is
\ba
E^{syn}_{p, th} \approx \cases{
1.5 \times 10^8 L_{52}^{-1/4}r_{12.5}^{1/2}\Gamma_{100}^{3/2}\, 
\mbox{GeV} & \mbox{(H)} \cr
2.5 \times 10^7 L_{52}^{-1/4}r_{11}^{1/2}\Gamma_{100}^{3/2}\, 
\mbox{GeV} & \mbox{(He);} }
\label{p-thresh-syn}
\ea
approximately an order of magnitude below the maximum proton energy
given in Eq. (\ref{maxproton-energy}). Due to the smaller
(sub-stellar) radii of these shocks and the different target photons,
the optical depths and threshold proton energies are respectively
higher and lower than those \cite{wb97} in $\gamma$-ray producing
shocks outside the stars.

Next, we consider $pp$ interactions in the internal shocks.  Typically
a fraction $\zeta_p < 1$ of the protons are Fermi accelerated in the
internal shocks of the GRB jet.  The remaining fraction $(1-\zeta_p)$
of cold protons provide targets for $pp$ interactions.  The mean total
cross-section for $pp$ interactions in the TeV-PeV energy range is $\<
\sigma_{pp} \> \approx 6 \times 10^{-26}$ cm$^2$.  The corresponding
optical depth in the jet frame, from Eq. (\ref{proton-jetdensity}), is
\ba 
\<\tau_{pp}\>_{jet} &=& (1-\zeta_p) n_{p, jet} \< \sigma_{pp} \>
r_{sh}/ \Gamma_{jet} \nonumber \\ &\approx & \cases{ 0.3 \,(1-\zeta_p)
L_{52} r_{12.5}^{-1} \Gamma_{100}^{-3} & \mbox{(H)} \cr 11
\,(1-\zeta_p) L_{52} r_{11}^{-1} \Gamma_{100}^{-3} & \mbox{(He);} }
\label{pp-sh-opt}
\ea
much smaller than $p\gamma$ optical depth.  At energies well below the
maximum proton energies, the competing effect of inverse Compton (IC)
scattering reduces the $pp$ scattering rate significantly, because of
the large synchrotron photon density in the internal shocks.

The cross-section for inverse Compton is $\sigma_{\rm IC} \approx
\sigma_{\rm Th} (m_e/m_p)^2$ in the non-relativistic Thomson limit.
The Thomson limit applies when $E_{\gamma}E_p/m_p^2 \ll 1$ in the
comoving frame.  Solving for the synchrotron photon energies
[Eq. (\ref{syn-phot-E})] in the $r_{12.5}$ and $r_{11}$ cases, the
Thomson limits are valid for $E_{p,jet}^{\rm IC} \lesssim 4 \times
10^6$ GeV and $7 \times 10^5$ GeV respectively in the jet frame.  The
same quantity in the observer's frame
\ba 
E_{p}^{\rm IC} \lesssim \cases{ 4 \times 10^8 L_{52}^{-1/4}
r_{12.5}^{1/2} \Gamma_{100}^{3/2} \, \mbox{GeV} & \mbox{(H)} \cr 7
\times 10^7 L_{52}^{-1/4} r_{11}^{1/2} \Gamma_{100}^{3/2} \,
\mbox{GeV} & \mbox{(He);} }
\label{proton-IC}
\ea
is higher than $E^{syn}_{p, th}$ [Eq. (\ref{p-thresh-syn})]. The ratio
of IC to $pp$ optical depth for the two internal shock radii is
\ba
\frac{\tau_{{\rm IC},jet}}{\<\tau_{pp}\>_{jet}} &=& \frac{n_{\gamma,
jet}^{syn} \sigma_{\rm Th} (m_e/m_p)^2} {n_{p, jet} \<\sigma_{pp}\>}
\nonumber \\ &\approx & \cases{ 14.6 \,L_{52}^{-1/4} r_{12.5}^{1/2}
\Gamma_{100}^{1/2} & \mbox{(H)}\cr 2.6 \,L_{52}^{-1/4} r_{11}^{1/2}
\Gamma_{100}^{1/2} & \mbox{(He).} }
\label{ICpp-ratio}
\ea
The proton flux is suppressed according to Eq. (\ref{ICpp-ratio})
below $E_{p}^{\rm IC}$ for $pp$ interactions in the internal shocks.
In the ultra-relativistic Klein-Nishina limit: $E_{\gamma}E_p/m_p^2
\gg 1$, the $pp$ optical depth becomes larger than the IC optical
depth with increasing proton energy.  However $p\gamma$ interactions
are dominant above $E^{syn}_{p, th}$ anyway.

Protons below $E^{syn}_{p, th}$, in the $r_{11}$ case, undergo $pp$
interactions in the internal shocks with reduced flux
[Eq. (\ref{ICpp-ratio})].  In the $r_{12.5}$ case, most protons below
$E^{syn}_{p, th}$ escape the internal shocks to interact with shocked
photons in the jet head and cold stellar protons.  We discuss these
interactions below.

The threshold proton energy at the $\Delta$ resonance for $p\gamma$
interactions with shocked photons in the jet head, from
Eq. (\ref{shgamma-temp}), is
\ba 
E_{p, th}^{sh} &=& 0.3 \Gamma_{h}/T_{h} \nonumber \\ &\approx &
3.2 \times 10^5 L_{52}^{-1/4} r_{12.5}^{1/2}\Gamma_{h,1.6}^{3/2}
\,\mbox{GeV} \;\; \mbox{(H)}
\label{p-thresh-sh} 
\ea
in the observer's frame.  The corresponding optical depth, in the jet
head frame, depends on the photon number density given in
Eq. (\ref{sh-gam-dist}) as
\ba 
\tau_{p\gamma}^{sh}(x) &=& l_{\gamma}/l_{p\gamma}(x) \approx
l_{\gamma} (\sigma_{p\gamma \ra \Delta}) n_{\gamma,h}\,
e^{-x/l_{\gamma}} \nonumber \\ &\approx & 1.46 \times 10^4 \,
e^{-x/l_{\gamma}} \nonumber \\ && \times \, L_{52}^{-1/4}
r_{12.5}^{1/2} \Gamma_{h,1.6}^{-1/2} \Gamma_{100} \;\; \mbox{(H)}.
\label{sh-gam-opt}
\ea
Since protons lose 20\% of their energy in each $p\gamma$ interaction,
it takes about 5 optical depths for a proton of energy $E_{p,
th}^{syn}$ in Eq. (\ref{p-thresh-syn}) to fall below $E_{p, th}^{sh}$.
Solving Eq. (\ref{sh-gam-opt}) for $x$ and $\tau_{p\gamma}^{sh}(x) =
5$, we have $x/l_{\gamma} \approx 8$.  The shocked photon density at
this distance, from Eq. (\ref{photon-headdensity}), is then
\ba 
n_{\gamma}^{sh} \approx 1.4 \times 10^{20} L_{52}^{3/4}
r_{12.5}^{-3/2} \Gamma_{h,1.6}^{-3/2} \, {\rm cm}^{-3} \;\; \mbox{(H)}
\label{sh-gam-dens}
\ea
in the jet head.

Protons below $E^{sh}_{p, th}$ in the $r_{12.5}$ case, interact with
cold stellar protons in the H envelope.  The number density of stellar
protons is $n_{p, {\rm H}} \approx 6 \times 10^{16}$ cm$^{-3}$.  The
corresponding $pp$ optical depth for cold stellar protons in the
stellar plasma frame is
\ba 
\< \tau_{pp} \>_* = \< \sigma_{pp} \> n_{p, {\rm H}}
(r_* - r_{jet}) \approx 2.5 \times 10^4 
\label{pp-star-opt}
\ea
which is very large and forces all protons to undergo $pp$
interactions.

\section{Neutrino production}
\label{sec:nuproduction}

High energy neutrinos are produced both in $p\gamma$ and $pp$
interactions dominantly through secondary pion ($\pi^{\pm}$) decays as
$\pi^{\pm} \ra \mu\nu_{\mu} \ra e\nu_e \nu_{\mu} {\bar \nu}_{\mu}$.
Each of the secondary leptons share roughly 1/4 of the initial pion
energy.

Because of the high $p\gamma$ optical depth in the internal shocks
[Eq. (\ref{pg-syn-opt})], pions and muons are produced in the
thermalized synchrotron photon bath.  The high magnetic field and
synchrotron photon density in the shocks may force high energy pions
and muons, from $p\gamma$ interactions, to lose their energies through
synchrotron radiation and inverse Compton scatterings before they
decay to neutrinos.  The same is true for $p\gamma$ interactions with
photons from the shocked plasma in the jet head for the $r_{12.5}$
case.  Neutrinos from $pp$ interactions taking place outside the jet
head in the unshocked stellar plasma, however, are not affected
because of the much smaller magnetic field and photon density in the
plasma.  We discuss synchrotron and inverse Compton losses in the GRB
jet below.

The synchrotron loss time for pions and muons in the jet frame can be
calculated using formulae similar to Eq. (\ref{sync-time}) for
protons.  Equating this time to the particle decay time in the jet
frame, $\tau_{\pi,jet}^{dec} = 2.6 \times 10^{-8} \gamma_{\pi, jet}$ s
and $\tau_{\mu,jet}^{dec} = 2.2 \times 10^{-6} \gamma_{\mu, jet}$ s,
we get the maximum synchrotron break energies, in the observer's
frame: $E_{\pi; \mu, jet}^{sb} = m_{\pi;\mu}\gamma_{\pi;\mu,jet}^{max}
\Gamma_{jet}$, as
\ba
E_{\pi}^{sb} \approx \cases{
1.3 \times 10^7 \xi_{-1}^{-1/2}L_{52}^{-1/2}r_{12.5}
\Gamma_{100} \, \mbox{GeV} & \mbox{(H)} \cr
4.2 \times 10^5 \xi_{-1}^{-1/2}L_{52}^{-1/2}r_{11}
\Gamma_{100}\, \mbox{GeV} & \mbox{(He)} }
\label{synloss-pi}
\ea
for pions and
\ba
E_{\mu}^{sb} \approx \cases{
7.0 \times 10^5  \xi_{-1}^{-1/2}L_{52}^{-1/2}r_{12.5}
\Gamma_{100} \, \mbox{GeV} & \mbox{(H)} \cr
2.2 \times 10^4 \xi_{-1}^{-1/2}L_{52}^{-1/2}r_{11}
\Gamma_{100} \, \mbox{GeV} & \mbox{(He)} }
\label{synloss-mu}
\ea
for muons.  

The inverse Compton (IC) losses are more severe for muons than for
pions.  The IC cooling time for a particle of mass $m_i$ in the
Thomson and Klein-Nishina (KN) limits is
\ba 
t^{\rm IC, Th}_{i,jet} &=& \frac{m_{i}}{\sigma_{\rm Th}c
\gamma_{i, jet} E_{\gamma, jet} n_{\gamma, jet}} \;\;; \gamma_{i, jet}
\ll \frac{m_{i}}{E_{\gamma, jet}} \;\; \nonumber \\ t^{\rm IC,
KN}_{i,jet} &=& \frac{ \gamma_{i, jet} (E_{\gamma,
jet}/m_{i})}{\sigma_{\rm Th}(m_e/m_{i})^2 c n_{\gamma, jet}} \;\;;
\gamma_{i, jet} \gg \frac{m_{i}}{E_{\gamma, jet}} \;\;
\label{IC-time} 
\ea
respectively in the jet frame.  The ratio of IC cooling time to the
particle decay time, for synchrotron photons from
Eqs. (\ref{syn-phot-E} \& \ref{syn-phot-density}), in the
Klein-Nishina limit, is
\ba
\frac{ t^{{\rm IC}, syn}_{\pi,jet} }{ \tau_{\pi,jet}^{dec} } \approx
\cases{ 2.6 \times 10^{-1} L_{52}^{-1/2}r_{12.5}\Gamma_{100} ; \cr \;\;
\gamma_{\pi,jet} \gtrsim 7 \times 10^5 & \mbox{(H)} \cr 8.9 \times
10^{-3} L_{52}^{-1/2}r_{11}\Gamma_{100} ; \cr \;\; \gamma_{\pi,jet}
\gtrsim 10^5 & \mbox{(He)} }
\label{icloss-syn-pi}
\ea
for pions and 
\ba
\frac{t^{{\rm IC}, syn}_{\mu,jet}}{\tau_{\mu,jet}^{dec}} \approx
\cases{ 2.3 \times 10^{-3} L_{52}^{-1/2}r_{12.5}\Gamma_{100} ; \cr
\;\; \gamma_{\mu,jet} \gtrsim 5 \times 10^5 & \mbox{(H)} \cr 7.9
\times 10^{-5} L_{52}^{-1/2}r_{11}\Gamma_{100} ; \cr \;\;
\gamma_{\mu,jet} \gtrsim 9 \times 10^4 & \mbox{(He)} }
\label{icloss-syn-mu}
\ea
for muons.  As a result, above $\gamma_{\pi;\mu, jet} \approx
m_{\pi;\mu}/E_{\gamma, jet}$, only $\approx$ 26\% of the pions
produced by $p\gamma$ interactions in the internal shocks, in the
$r_{12.5}$ case, decay to $\nu_{\mu}$.  Production of $\nu_e$ and
${\bar \nu}_{\mu}$, from $\mu$-decay, in the $r_{12.5}$ case is
suppressed by IC losses.  All flavors of neutrino production in the
internal shocks from $p\gamma$ interactions is suppressed in the
$r_{11}$ case by IC losses.  Below $\gamma_{\pi;\mu, jet} \approx
m_{\pi;\mu}/E_{\gamma, jet}$ neutrino production is suppressed as
$E_{\nu}^{-2}$ due to IC losses in the Thomson limit.  Similarly, for
$pp$ interactions in the internal shocks, neutrino production is also
suppressed by IC losses.

Neutrino production with the shocked photons in the jet head, in the
$r_{12.5}$ case, is also affected by inverse Compton losses.  The
decay length $c\tau_{\pi;\mu}$ in the shocked plasma frame exceeds the
shocked photon penetration depth [Eq. (\ref{photon-depth})] for
$\gamma_{\pi} \gtrsim 9 \times 10^4$ and $\gamma_{\mu} \gtrsim 10^3$.
As a result pions and muons produced in the $p\gamma$ interactions
``see'' an increasing photon density according to
Eq. (\ref{sh-gam-dist}).  For simplicity we compute the ratios of IC
loss times to the decay times for pions and muons, in the
Klein-Nishina limit: $\gamma_{\pi} \gtrsim 9 \times 10^4$ and
$\gamma_{\mu} \gtrsim 6 \times 10^4$, as
\ba 
t^{{\rm IC}, sh}_{\pi}/\tau_{\pi}^{dec} &\approx & 3.6 \times 10^{-3}
L_{52}^{-1/2}r_{12.5}\Gamma_{h,1.6}  \;\; \mbox{(H)} \nonumber \\ 
t^{{\rm IC}, sh}_{\mu}/\tau_{\mu}^{dec} &\approx & 3.2 \times 10^{-5}
L_{52}^{-1/2}r_{12.5}\Gamma_{h,1.6}  \;\; \mbox{(H)}
\label{icloss-sh}
\ea
using Eqs. (\ref{shgamma-temp} \& \ref{photon-headdensity}).  As a
result, all flavors of neutrino production is suppressed by IC losses
in the jet head.

\section{Neutrino flux calculation}
\label{sec:nuflux}

The proton energy distribution in the sub-stellar internal shocks, in
the observer frame, is given by
\ba 
\frac{d^2N}{dE_p dt} = \frac{\zeta_p \vareps_{\rm op}
L_{\gamma}^{iso}}{E_p^2} \approx 6 \times 10^{54} \frac{\zeta_p
\vareps_{\rm op} L_{52}}{E_p^2} \, {\rm GeV}^{-1} {\rm s}^{-1},
\label{proton-spectrum} 
\ea
where $\zeta_p$ is the injection fraction of protons into the
acceleration process and $\vareps_{\rm op}$ is the fraction of jet
energy dissipated in sub-surface shocks (whose order of magnitude is
$\sim 1/2$).  The neutrino flux, the same for $\nu_{\mu}$, $\nu_e$ and
${\bar \nu}_{\mu}$ in the $pp$ and $p\gamma$ interactions, from a
single GRB buried jet at a distance $D$ is \cite{Razzaque:2002kb}
\ba
\Phi_{\nu} &=& d^2N/dE_\nu dt = 1/4\pi D^2 \nonumber \\ & \times &
\left\{ \begin{array}{ll} \int f_{pp}\, M_{\nu} (E_p) \frac{d^2N}{dE_p
dt} dE_p &; E_p \lesssim E_{p}^{\rm th} \\ (f_{\pi}/4)\,
\frac{d^2N}{dE_p dt} &; E_p > E_{p}^{th}
\end{array} \right.
\ea
where $f_{pp} = {\rm min} (1,\<\tau_{pp}\>)$ and $f_{\pi} = {\rm min}
(1,\tau_{p\gamma})$ from Eqs. (\ref{pp-sh-opt} \& \ref{pg-syn-opt})
respectively.  We have used the {\small PYTHIA} 6.2 event generator
\cite{pythia}, widely used in high energy particle physics, to
simulate $pp$ interactions. For our problem, the neutrino
multiplicity, through pion decay, in the $pp$ interactions can be
written as \cite{Razzaque:2002kb}
\ba
M_{\nu}(E_p) &=& \frac{7}{4} \left(
\frac{E_{\nu}}{\rm GeV} \right)^{-1} \left[ \frac{1}{2}{\rm ln}\left(
\frac{10^{11}\, {\rm GeV}}{E_p} \right) \right]^{-1} \nonumber \\
&\times & \Theta \left( \frac{1}{4} \frac{m_{\pi}}{\rm GeV}
\gamma_{\rm CM} \leq \frac{E_{\nu}}{\rm GeV} \leq \frac{1}{4
}\frac{E_p}{\rm GeV} \right)
\label{pp-numult}
\ea
for each flavor of neutrino.  However, observed neutrino fluxes, for
different flavors, are greatly affected by the environments at which
they are produced.  We discuss our flux calculation method next.

\begin{figure}[htb]
\centerline{\epsfxsize=3.4in \epsfbox{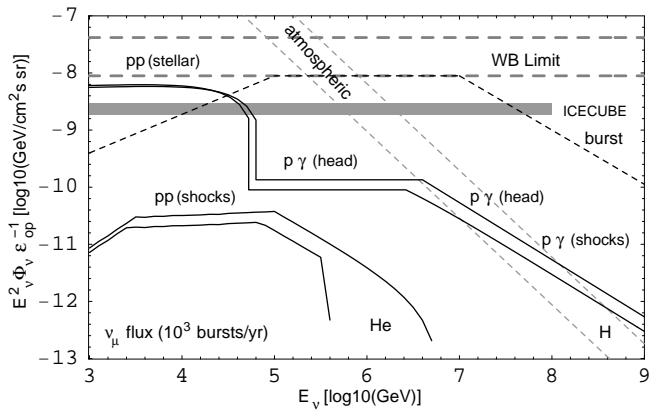}}
\caption{Diffuse muon-neutrino flux $E_{\nu}^2 \Phi_\nu \vareps_{\rm
op}^{-1}$, shown as solid lines, from sub-stellar jet shocks in two
GRB progenitor models, H ($r_{12.5}$) and He ($r_{11}$), each for two
sets of shock/jet radii and overlying envelope masses (with similar
curves for $\nu_{\mu}$, $\nu_e$ and $\nu_{\tau}$).  These neutrinos
arrive as precursors (10-100 s before) of $\gamma$-ray bright
(electromagnetically detectable) bursts, or as the sole signal in the
case of $\gamma$-ray dark bursts.  Also shown is the diffuse neutrino
flux arriving simultaneously with the $\gamma$-rays from shocks
outside the stellar surface in observed GRB (dark short-dashed curve);
the Waxman-Bahcall (WB) diffuse cosmic ray bound (light long-dashed
curves); and the atmospheric neutrino flux (light short-dashed
curves). For a hypothetical 100:1 ratio of $\gamma$-ray dark (in which
the jets do not emerge) to $\gamma$-ray bright collapses, the neutrino
fluxes would be 100 times higher than those plotted here.}
\label{fig:difflux}
\end{figure}

In the $r_{11}$ (He) case, all flavors of neutrinos produced from
$p\gamma$ and $pp$ interactions are heavily suppressed by inverse
Compton losses both in the Thomson and Klein-Nishina limits
[Eqs. (\ref{icloss-syn-pi} \& \ref{icloss-syn-mu})].  We have
calculated the muon neutrino flux (see Fig. \ref{fig:difflux}) from
$pp$ interactions below $E^{syn}_{p, th}$ [Eq. (\ref{p-thresh-syn})],
in the internal shocks, using $\sim$ 40\% of the proton flux
[Eq. (\ref{proton-spectrum})] according to Eq. (\ref{ICpp-ratio}).
Above $E^{syn}_{p, th}$ (corresponding to $E_\nu \sim 10^{6.5}$ GeV)
we used the full proton flux to calculate neutrino flux from $p\gamma$
interactions (not shown in the figure).

The $r_{12.5}$ (H) case is more complicated.  Below $E^{sh}_{p, th}$
(Eq. [\ref{p-thresh-sh}], corresponding to $E_\nu \sim 10^{4.5}$ GeV),
we calculate neutrino flux through $pp$ interactions with cold stellar
protons in the H envelope.  We used $\sim$ 70\% of the proton flux
[Eq. (\ref{proton-spectrum})] escaping the internal shocks according
to Eq. (\ref{pp-sh-opt}) for this calculation.  Note that neutrinos
thus produced in the H envelope are not suppressed by inverse Compton
losses.  Above $E^{sh}_{p, th}$ and below $E^{syn}_{p, th}$
[Eq. (\ref{p-thresh-syn})] $\nu_{\mu}$ production from $p\gamma$
interactions, at the jet head, is suppressed by IC losses
[Eq. (\ref{icloss-sh})].  Similar IC suppression
[Eqs. (\ref{icloss-syn-pi} \& \ref{icloss-syn-mu})] applies for
$\nu$'s from $p\gamma$, in the internal shocks, above $E^{syn}_{p,
th}$.  An extra reduction of the flux happens above $E_{\pi}^{sb}$
[Eq. \ref{synloss-pi}] due to synchrotron radiation by pions.

The diffuse $\nu_\mu$ flux levels are shown in Fig. \ref{fig:difflux}
as solid curves for the H (upper two curves) and He (lower two curves)
star models. The nominal set of jet and shock radii discussed in \S
\ref{sec:jetmodel} and above ($\alpha \sim 1$, H: $r_{12.5}$ and He:
$r_{11}$) give the uppermost solid curve of each set. For comparison,
we also show the flux for a different choice of jet and shock radii,
plotted as the lower of each set of solid curves; these latter have
$r_{jet,{\rm H}} \sim 10^{12.3}$ cm, $r_{sh,{\rm H}} \sim 10^{12}$ cm
in the H case, and $r_{jet,{\rm He}} \sim 10^{10.8}$ cm and
$r_{sh,{\rm He}} \sim 10^{10.5}$ cm in the He case respectively.  Note
that, because of a $pp$ component, the neutrino flux we calculated is
higher in the $<100$ TeV range compared to optically thin shocks
\cite{wb97, wb99}.  As a result, one gets more events in a neutrino
detector than one would from optically thin shocks \cite{dermer03}.

The effects of absorption by $\nu N$ interactions in the overlying
stellar envelope at the highest energies plotted in
Fig. \ref{fig:difflux} are calculated using the cross sections
\cite{nucross} and the overlying grammages above the jet head. The
absorption is negligible in the general case ($\alpha \sim 1$,
uppermost solid curve in each model), which have roughly $\sim 0.1
M_\odot$ (over $4\pi$) of overlying envelope. For jet and shock radii
occuring further in, as in the lower of each set of curves,
the overlying envelope material that we took is more massive, $\sim 1 
M_\odot$. In this case the absorption effects can become noticeable, 
especially in He models, and this might serve as a diagnostic of the 
sub-surface jet depth (or the overlying material). For a solar mass 
above the jet head (distributed over $4\pi$), the optical depth for 
$\nu N$ interactions becomes larger than unity at energies 
$E_{\nu} \gtrsim 2.5 \times 10^{11}$ GeV in the H case and 
$E_{\nu} \gtrsim 2.5 \times 10^{5}$ GeV in the He case (seen as a 
cut-off in the lower set of He curves). Typically 80\% of the 
$\nu$-energy is transferred to the secondary lepton: $e$ or $\mu$. 
These $\mu$'s further decay to $\nu$'s, however, they are subject to 
severe synchrotron and inverse Compton losses and do not contribute 
substantially to lower energy flux.

The diffuse fluxes are based on a conservative estimate of the source
density from the observation that $\sim 10^3$ GRB are known to occur
per year over the entire sky, based on $\gamma$-ray detections. If
these are attributed to massive stellar collapses, each GRB should be
preceded by neutrino percursor events, starting tens of seconds
(typically $\sim$ 30 s \cite{Waxman:2003uu}) before the $\gamma$-rays.
The total $\nu_\mu$ diffuse flux from all of these ($\gamma$-ray
quasi-coincident) events is given by the solid curves in
Fig. \ref{fig:difflux}. (Considering mass-mixing along the way to
Earth from $z\sim 1$, the ratio of flavors as observed from earth
should become unity). The number of neutrino bursts and
correspondingly the diffuse neutrino flux may be up to a factor $100
q$ higher than what is shown in Fig. \ref{fig:difflux}, if the ratio
of the number of buried jets which do not emerge ($\gamma$-ray dark
bursts) to that of those which do emerge is $100q$, where $q\lesssim
1$.

The number of muon neutrino events, in a km$^2$ detector (IceCube
\cite{icecube} e.g.), from a single GRB at redshift $z=1$ with buried
jet of duration $\Delta t \approx 30$ s is 0.6 and 0.003 in the
$r_{12.5}$ (H) and in the $r_{11}$ (He) cases, respectively, in the
1-100 TeV energy range.  The number of events can be significantly
larger for brighter ($L_{\gamma}^{iso} > 10^{52}$ ergs/s) or nearer
($z<1$) bursts, or also if they are of longer duration ($\Delta t >
30$ s).

\section{Implications}
\label{sec:implications}

The main feature of the high energy neutrino spectra of massive
stellar collapse buried jet models related to GRB (and also of models
involving a precursor supernova, Refs. \cite{Razzaque:2002kb, gg02}),
is that a ``thicker target'' of nucleons and photons is available to
give higher fluxes of $\sim$ 1-100 TeV neutrinos, compared to those
expected from internal or external shocks which occur outside the
stellar progenitor \cite{wb97, wb99}.

The neutrino flux expected in association with electromagnetically
detected GRB, shown in Fig. \ref{fig:difflux}, precedes the
$\gamma$-rays by about 10-100 seconds, approximately the time taken by
the jet to emerge from the collapsing core and the stellar envelope.
Neutrinos coming from both $p\gamma$ and $pp$ interactions in the
sub-stellar jet are heavily suppressed by inverse Compton scatterings
due to large photon density.  The precursor neutrinos in the 1-100 TeV
range from a single $\gamma$-ray bright event, coming from $pp$
interactions in the H envelope ($r_{12.5}$ case) can be detected with
ICECUBE.  The number of events from a single burst at a distance
$D=10^{28.2}$ cm ($z\sim 1$) and duration $\Delta t=100$ s would be
$\sim$ 4 in the $r_{12.5}$ H-envelope star case, compared to
$\lesssim$ 0.04 in the $r_{11}$ He star case. (Rare single bursts
occuring about once per year at $D \sim 10^{27.2}$ cm could yield
$\sim 10^2$ higher event rates per burst). Thus, detection of many
bursts with 1-100 TeV neutrino precursors in correlation with a
subsequent $\gamma$-ray detectable GRB would be a strong indication
for an H progenitor.  A rare detection of a weaker 100 TeV neutrino
precursor would favor a He star progenitor.

The number of stars more massive than about 30 solar masses out to
$z\sim 1$ undergoing a core collapse that leads to a black hole is
about $10^7$/year corresponding to $\sim 3 \times 10^{-2}$ yr$^{-1}$
galaxy$^{-1}$ of type II supernovae rate. An interesting possibility
\cite{mw01} is that many or perhaps all of these lead to jets as in
GRB, in which case most of these jets do not break through the stellar
surface, and hence are not observed in $\gamma$-rays (at most $\sim
10^3$/year can be observed as GRB). If the mean jet solid angle
subtends 1\% of the sky, this means $\lesssim 10^5$/year such
$\gamma$-ray dark (choked) bursts whose jets are directed at
earth. These would be detectable only through their neutrino emission,
which should be the same per burst as for the $\gamma$-ray bright
bursts, but they would be up to 100 times more numerous. The diffuse
$\nu_\mu$ flux from these $\gamma$-ray dark collapses would be similar
to that of the $\gamma$-ray bright solid neutrino curves of
Fig. \ref{fig:difflux}, but up to 100 times higher.

The diffuse neutrino signals from both $\gamma$-ray bright and dark
collapses should be detectable with IceCube, as can be seen from the
sensitivity curves in Fig. \ref{fig:difflux}. For the $\gamma$-ray
bright case plotted, the TeV-PeV flux detection is aided by the
positional and temporal coincidences with electromagnetic detections.
The typical angular resolution of planned neutrino telescopes at TeV
energies \cite{icecube} should be $\theta\lesssim 1^o$.  For the
$\gamma$-ray dark cases, the diffuse flux for a total number of
collapses $\lesssim 10^5$/year out to $z\sim 1$ would be sufficiently
above the atmospheric neutrino background, especially in the PeV
range, that detection appears possible even without coincident photon
flashes.  (The $\gamma$-ray dark collapses could, however, be possibly
associated with faint supernova-like optical/IR events, weeks to
months later, if the stellar envelope is ejected).

The non-detection of a diffuse signal would be significant only if we
have direct evidence for neutrino production in GRBs.  This may come
from a detection of either the GRB/afterglow neutrinos, or from a
detection of the stellar $pp$ neutrinos in the H case.  In this case,
an upper limit on the diffuse flux can serve to constrain the rate of
``dark" bursts and the progenitor nature.  If there is no direct
evidence for neutrino production, the non-detection of a diffuse
signal may imply no protons in the GRB jet and/or no proton
acceleration.

The flux may exceed the WB bound \cite{wb99} if opaque collisions
occur where the $pp$ ($p\gamma$) optical depth is very high, which
implies that the nucleons do not escape. In this case, the proton
energy generation rate of these ``hidden sources'' may exceed the
energy generation rate inferred from cosmic-ray observations, and
hence the neutrino flux may exceed the WB bound.

The neutrino flux level and spectra in detected GRBs would serve as a
diagnostic of the outer dimensions of the stellar progenitor at the
time the explosion occured, providing useful constraints on the
identity and evolutionary scenarios of the progenitors, as well as
providing clues and a consistency check for possible association with
supernova-like objects. The measurement of such neutrino bursts from
$\gamma$-ray dark collapses would provide useful constraints on the
total rate of massive stellar collapses at high redshifts, a quantity
of significant interest for cosmological reionization, structure
formation and intergalactic metal enrichment scenarios.

\noindent
{\it Acknowledgements-} This work was supported in part by NSF
AST0098416.


\begin{thebibliography}{}
\def\bitm{\bibitem}

\bitm{woo03} S.~E.~Woosley, W.~Zhang \& A.~MacFadyen, Astrophys.\ J.\
in press [arXiv:astro-ph/0207436] 

\bitm{mesz02} P.~\Mesz, Ann.\ Rev.\ Astr.\ Astrophys.\ {\bf 40}, 137
(Annual Reviews: Palo Alto) (2002).

\bitm{wax95} E.~Waxman, Phys.\ Rev.\ Lett.\ {\bf 75}, 386 (1995).

\bitm{wb97} E.~Waxman \& J.~N.~Bahcall, Phys.\ Rev.\ Lett.\ {\bf 78},
2292 (1997).

\bibitem{mw01} P.~\Mesz and E.~Waxman, Phys.\ Rev.\ Lett.\ {\bf 87},
171102 (2001) [arXiv:astro-ph/0103275].

\bitm{belob00} A. Beloborodov, B. Stern, R. Svensson, 
ApJ, 535, 158 (2000); E. Woods, A. Loeb, ApJ 453, 583 (1995)

\bibitem{dermer02} C.D.~Dermer, Astrophys.\ J.\ 574:65, 2002.

\bibitem{Waxman:2003uu} E.~Waxman and P.~\Mesz, Astrophys.\ J.\ {\bf
584}, 390 (2003).

\bitm{wb99} E.~Waxman \& J.~N.~Bahcall, Phys.\ Rev.\ {\bf D} 59, 3002
(1999).

\bibitem{Razzaque:2002kb} S.~Razzaque, P.~\Mesz and E.~Waxman,
arXiv:astro-ph/0212536.

\bibitem{pythia} T.~Sjostrand {\it et al.}, Comp.\ Phys.\ Commun.\
{\bf 135}, 238 (2001); T.~Sjostrand {\it et al.},
arXiv:hep-ph/0108264.

\bitm{dermer03} C.D.~Dermer and A.M.~Atoyan, astro-ph/0301030.

\bibitem{nucross} Yu.~Andreev {\it et al.}, Phys.\ Lett.\ B {\bf 84},
247 (1979); D.~McKay and J.~Ralston Phys.\ Lett.\ B {\bf 167}, 103
(1986); G.~Frichter {\it et al.}, Phys.\ Rev.\ Lett.\ {\bf 74} (1995),
{\it Erratum-ibid.} {\bf 77}, 4107, (1996); R.~Gandhi {\it et al.},
Phys.\ Rev.\ D {\bf 58}, 093009 (1998).

\bitm{icecube} IceCube Collaboration, J.~Ahrens {\it et al.}
astro-ph/0305196.

\bibitem{gg02} J.~Granot and D.~Guetta, arXiv:astro-ph/0211433;
D.~Guetta and J.~Granot, arXiv:astro-ph/0212045.

\end{thebibliography}
\end{document}